\documentclass[]{article}

 \usepackage[]{graphicx}
 \usepackage{epsfig}
 \usepackage{amsmath, amstext, amsfonts}
 \usepackage{amssymb}
 \usepackage{multicol}
 \usepackage[latin1]{inputenc}

\usepackage{anysize} 
\marginsize{2cm}{2.5cm}{1.5cm}{2cm}

 \title{A Simple Electronic Circuit Realization of the Tent Map}

 \date{}

\begin{document}

 \maketitle
 \begin{center}{\sc 
 I. Campos-Cantón$^{1}$,  E. Campos-Cantón$^{2}$, J. S. Murguía$^{2}$ and H. C. Rosu$^{3}$
 \footnotesize \vskip 4ex
   $^1$Fac. de Ciencias, $^2$Departamento de Físico Matemáticas, \\
   {\em Universidad Aut\'{o}noma de San Luis Potos\'{\i}},\\
       {\em Alvaro Obreg\'on 64, 78000, San Luis Potos\'{\i}, SLP,
       M\'{e}xico}\\
   $^3$Divisi\'on de Materiales Avanzados,\\ {\em Instituto Potosino de Investigaci\'on Cient\'{\i}fica y Tecnol\'ogica}\\
   {\em Camino a la presa San Jos\'e 2055, 78216, San Luis Potos\'{\i}, SLP,
       M\'{e}xico}\\
       {\footnotesize icampos@galia.fc.uaslp.mx, ecamp@uaslp.mx, ondeleto@uaslp.mx, hcr@ipicyt.edu.mx}}
 \end{center} 

 \begin{abstract}
  \footnotesize
   \noindent
   We present a very simple electronic implementation of the tent map, one of the best-known discrete dynamical systems.
   This is achieved by using integrated circuits and passive elements only. The experimental behavior of the tent map electronic circuit
   is compared with its numerical simulation counterpart. We find that the
   electronic circuit presents fixed points, periodicity, period doubling,
   chaos and intermittency that match with high accuracy the corresponding theoretical values.\\
\\
PACS:84.30.-r; 05.45.-a; 02.30.Oz \hfill TMap-fin1.tex

\noindent KEYWORDS: Electronic circuits, nonlinear dynamics and
chaos, bifurcation theory.\hfill arXiv: 0807.3375
 \end{abstract}

 \begin{center}
 {\tt Chaos, Solitons \& Fractals 42 (2009) 12-16}
 \end{center}




\section{Introduction}   \label{S_Int}

\begin{figure}
 \centering
 \includegraphics[width= 7.5 cm, height=5.5 cm]{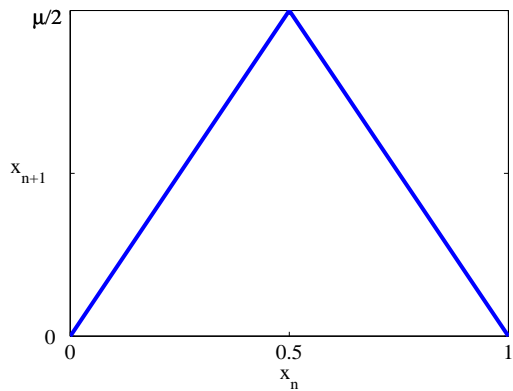}
\caption{\sl Plot of the tent map function.}
 \label{fig tent map}
 \end{figure}

Discrete-time nonlinear dynamical systems are generally described as
iterative maps $f : \Re^k \rightarrow \Re^k$ given by their state
equation

\begin{equation}\label{ecite}
 \textbf{x}_{n+1} = f(\textbf{x}_n), ~~~  n = 0, 1, 2, . . . ~,
 \end{equation}
\\

\noindent where $\textbf{x}_0$ is the initial state, $k$ is the
dimensionality of the state space, $\textbf{x}_n \in \Re^k$ is the
state of the system at time $n$, and $\textbf{x}_{n+1}$ denotes the
next state. The interpretation of the state vector depends on the
context. For example, in population biology $\textbf{x}_n$ is
usually the population size in generation $n$, in epidemiology it is
the fraction of the population infected at time $n$, whereas in
economics it can be the price per unit at time $n$ for a certain
commercial product. Repeated iteration of $f$ gives a sequence of
points ${\{\textbf{x}_n\}}^{\infty}_{n=0} $ that is known as an
orbit. Clearly, equation \eqref{ecite} is a difference equation. In the words
of R.M. May \cite{May76}, {\em such equations, even though simple
and deterministic, can exhibit a surprising array of dynamical
behaviour, from stable points, to a bifurcating hierarchy of stable
cycles, to apparently random fluctuations}. The tent map is one of
the simplest iterated functions and, either alone or in more general forms, has been the subject of interesting papers published in this journal,
see e.g., \cite{csf-t1,csf-t2,csf-t3}. It has the
shape of a tent as is shown in Fig.~\ref{fig tent map}. 
It takes a point $x_n$ on the real line and maps it to another point
given by the following equation

\begin{equation}\label{ectent}
 x_{n+1} =\left\{
  \begin{matrix} \mu x_n & \mathrm{for}~~ x_n < \frac{1}{2} \\ \\ \mu (1-x_n) & \mathrm{for}~~ \frac{1}{2} \leq x_n ,
 \end{matrix}\right.
 \end{equation}
\\
where $x_n \in [0, 1]$, and $\mu \in [1, 2]$ is a bifurcation
parameter that controls the properties of the tent map. Many of the basic properties of the tent map
can be found in the book of Elaydi \cite{elaydi} on discrete chaos.\\

The tent map is a very simple model for studying a variety of
nonlinear phenomena.
The nonlinear dynamics of the tent map has found applications in as
different areas as biophysics, meteorology, hydrodynamics, chemical
engineering, optics, cryptology, and communications. For example, in
\cite{hasler97} the tent map is used to illustrate the
synchronization and/or non-synchronization of chaos, which raised
considerable interest in finding simple circuits that exhibit
nonlinear phenomena. Murali {\em et al} \cite{murali03} provided a
proof of principle experiment of the capability of chaotic systems
for universal computing.

\begin{figure}[h!]
 \centering
 \includegraphics[width= 6.5 cm, height=5.0 cm]{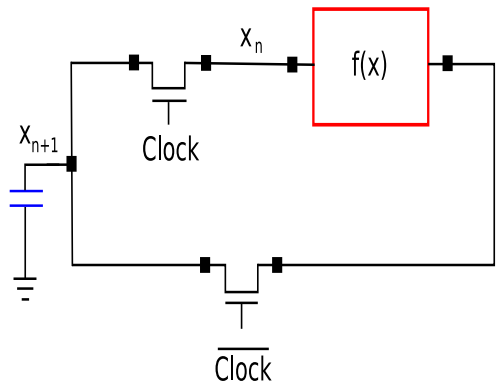}
\caption{\sl A typical block diagram of a map.}
 \label{map chaos}
 \end{figure}

In general, any map can be electronically designed. Following
Tanaka {\em et al.} \cite{tan00}, a typical circuit diagram of a
chaotic one-dimensional map with its iterative operation is shown in
Fig.~\ref{map chaos}. In this paper, we present one of the simplest
electronic implementation of the tent map, which at the same
time is a good engineering model of the corresponding mathematical
system. Through the variation of the tent map control parameter
$\mu$, one can examine the bifurcation diagram of the realized
system and we were able to reproduce the theoretical diagram with
high accuracy.

\section{Electronic implementation of the tent map}\label{S_ei}

\begin{figure}
 \centering
 \includegraphics[width= 13.5 cm, height=4.5 cm]{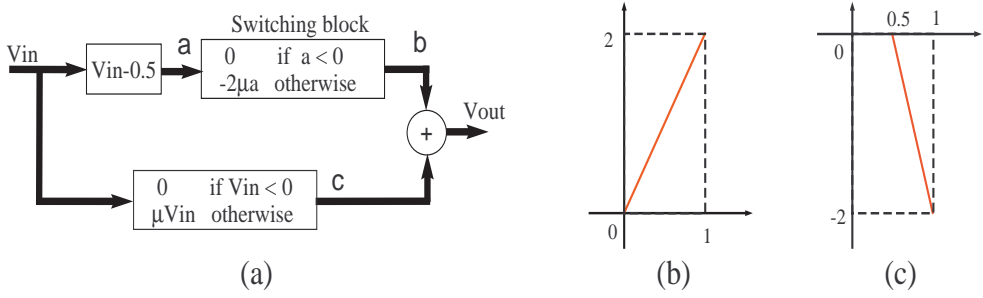}
\caption{\sl ({\rm a}) Block diagram of the tent map used to construct the electronic circuit. ({\rm b}) Response of the lower branch of the block diagram. ({\rm c}) Response of the upper branch of the block diagram.}
 \label{bloques}
 \end{figure}
In several implementations of this kind of circuits
\cite{Madhekar_06, corron_07} analog multipliers have been employed
with a normalization of the signal by a factor of about ten. This
normalization was necessary because of the physical restrictions in
the analog multiplier. The starting point is a block diagram of the
tent map that is shown in Fig.~\ref{bloques}. Typically, these
circuits contain several operational amplifiers, which perform
linear operations (e.g., integration and summation), as well as a
couple of integrated circuits that perform the nonlinear operations
(i.e., multiplication). In general, a large number of active
components make it difficult to directly extrapolate these designs
to high frequencies. Another approach is to use a digital signal
processor and digital-to-analog converters. Here, we describe a new
circuit that contains active components, speeds of radio
frequencies, and capable of reproducing the transition from steady
state to chaos as observed in the tent map equation when the
bifurcation parameter is varied.\

\begin{figure}
 \centering
 \includegraphics[width= 10.5 cm, height=7.5 cm]{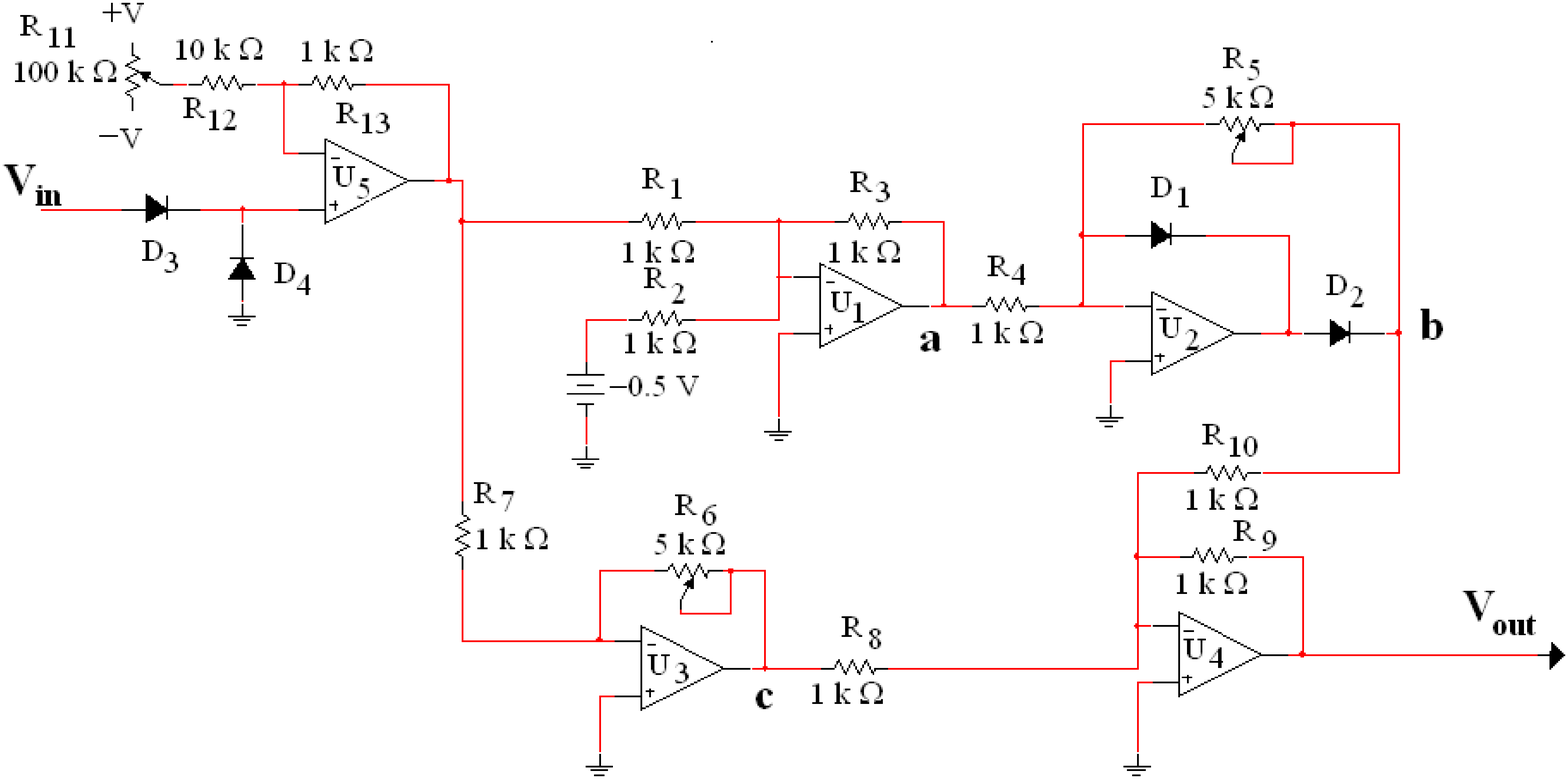}
\caption{\sl Schematic diagram of the tent map electronic circuit.}
 \label{esquematico}
 \end{figure}

We now introduce a designed circuit of the tent map based on Fig.~\ref{bloques}.
The flow diagram of the tent map used to construct the electronic circuit is
shown in Fig.~\ref{bloques} (a). The behavior of the tent map is based on two straight lines given by $l_1: \mu V_{in}$
and $l_2: -2\mu V_{in}+1$ with domains $[0,\infty)$ and $[0.5, \infty)$, respectively. The output $V_{out}$ is given by $l_1$ when $V_{in}$ belongs to the interval [0V, 0.5V) and by $l_1+l_2$ when $V_{in}$ is in the interval [0.5V, 1V).
The responses of the lower and upper branches are shown in Figs.~\ref{bloques} (b) and (c), respectively. This simple approach allows for
the changing of the slope from $\mu$ to $-\mu$. One can think of the system as having two weak points, $V_{in} <0$ and $V_{in} >1$.
However, the response of the circuit is zero for these inputs. In the absence of noise the tent map circuit can remain in one of the fixed points,
but in the real world of analog electronic components there always exists some noise that generates the dynamics in the circuit.
The schematic diagram of the tent map circuit is shown in Fig.~\ref{esquematico}, which consists of five operational
amplifiers (from $U_1$ to $U_5$), four diodes ($D_1$ - $D_4$), thirteen resistors (from R$_1$ to
R$_{13}$), and a dc voltage source (Vdc). The simplicity of this circuit
is due to the fact that the linear mathematical operation of
commutation is performed by the operational amplifiers in the
switching block, as is shown in Fig.~\ref{bloques}.\

Assuming ideal performance from all components, the circuit in
Fig.~\ref{esquematico} is modeled by the following equation

\begin{equation}\label{ecelect}
 V_{out} = \frac{{\rm R}_9{\rm R}_6}{{\rm R}_8{\rm R}_7}V_{in}-\left\{
\begin{array}{ll}
    0, & \hbox{for  $V_{in} < \displaystyle\frac{{\rm R}_1}{2{\rm R}_2}$}~, \\\\
   \displaystyle\frac{{\rm R}_9{\rm R}_5}{{\rm R}_{10}{\rm R}_4}\left(\frac{{\rm R}_3V_{in}}
  {{\rm R}_1}-\frac{{\rm R}_3}{2{\rm R}_2}\right) , & \hbox{for $V_{in} \geq \displaystyle\frac{{\rm R}_1}{2{\rm R}_2}$~,}
  \end{array}\right.
 \end{equation}\\
where $V_{in}$ and $V_{out}$ are the input and output voltages of
the tent map electronic circuit, respectively. It is worth noting
that the switching block shown in Fig.~\ref{bloques} is
realized through the \textbf{b} node,\\

\begin{equation}\label{ecb}
V_{{\rm \bf  b}} = \left\{
  \begin{array}{ll} 0, & \hbox{for~~ $V_{in} < \displaystyle\frac{{\rm R}_1}{2{\rm R}_2}$}~, \\ \\
                    \displaystyle\frac{{\rm R}_5}{{\rm R}_4}\left(\frac{{\rm R}_3V_{in}}{{\rm R}_1}-\frac{{\rm R}_3}{2{\rm R}_2}\right), & \hbox{for~~ $V_{in} \geq \displaystyle\frac{{\rm R}_1}{2{\rm R}_2}$}~.
 \end{array}\right.
 \end{equation}
\\

\begin{table}
  \centering
\begin{tabular}{|l|l|}
  \hline
  Device & Value \\\hline
  R$_{1,2,3,4,7,8,9,10,12,13}$  & 1 $k\Omega$ resistor \\\hline
  R$_{5,6}$               & 5 $k\Omega$ potentiometer\\\hline
  R$_{11}$               & 100 $k\Omega$ potentiometer\\\hline
  $D_{1,...,4}$               & 1n1419 diode \\ \hline
  $U_{1,...,5}$           & LM324 op. amp. \\\hline
\end{tabular}
\caption{The values of the electronic components employed in the construction of the
tent map electronic circuit.}\label{tabla1}
\end{table}

Thus, Eq.~\eqref{ecelect} is equivalent to Eq.~\eqref{ectent} for
the values of the components given in Table~\ref{tabla1}, and replacing
$V_{in}$ and $V_{out}$ for $x_n$ and $x_{n+1}$, respectively. In
fact, this set of values is not unique because Eq.~\eqref{ecelect}
contains several parameters. Thus, a circuit designer has the
freedom to choose the particular components that satisfy other
design constraints in a particular application. Despite of parasitic
reactance, finite bandwidth of active components, and other
experimental perturbations, the presented electronic circuit
displays closely the behavior of the mathematical model given by
Eq.~\eqref{ectent}. We implemented this design on a printed circuit
board (PCB) manufactured in our laboratory. In the experimental
circuit we used the LM324 operational amplifiers supplied with a
power source at $\pm 15$V and soldered directly to the PCB without a
socket. The voltage Vdc was supplied by a variable dc supply with an
 output range of $0-15$V. In order to have an iterative operation, see Fig. \ref{map chaos},
 this circuit considered a microcontroller PIC16F877A of Microchip,
 and a D/A converter DAC0800 of National Semiconductors
 with a processing time of 100 $\mu$s
 between voltage samples. Obviously, there are different ways to
 perform this iterative operation, but this is a matter that depends of the
 designer and the application.

\begin{figure}[h!]
 \centering
 \includegraphics[width= 12. cm, height=6. cm] {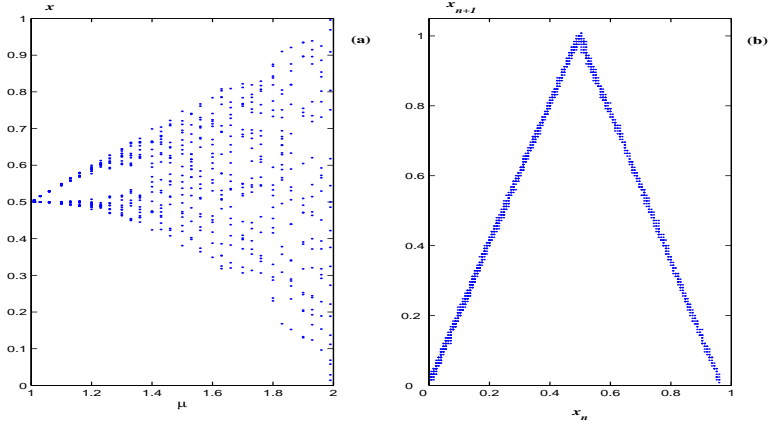}
\caption{\sl Experimental bifurcation diagram for the tent map.}
 \label{figdatos}
 \end{figure}

The value of the bifurcation parameter $\mu$ can be fixed at certain
values by simply adjusting the potentiometers $R_5$ and $R_6$
located in the operational amplifiers $U_2$ and $U_3$. The
relationship between the resistors R$_5$ and R$_6$ with the value of
$\mu$ is given by equation (\ref{vout-bis}), i.e.,
$\mu = {\rm R}_5/2{\rm k}\Omega={\rm R}_6/1{\rm k}\Omega$.

\begin{equation}\label{vout-bis}
V_{out} = \left\{
  \begin{array}{ll} \displaystyle\frac{{\rm R}_6}{1{\rm k}\Omega}V_{in}, & \hbox{for~~ $V_{in} < \displaystyle\frac{1}{2}$}~, \\ \\
                    \displaystyle\frac{{\rm R}_5}{2{\rm k}\Omega}\left(1-\left(2-\frac{2{\rm R}_6}{{\rm R}_5}\right)\right), & \hbox{for~~ $V_{in} \geq \displaystyle\frac{1}{2}$}~.
 \end{array}\right.
 \end{equation}
\\

\begin{figure}[h!]
 \centering
 \includegraphics[width= 9. cm, height=5.5 cm] {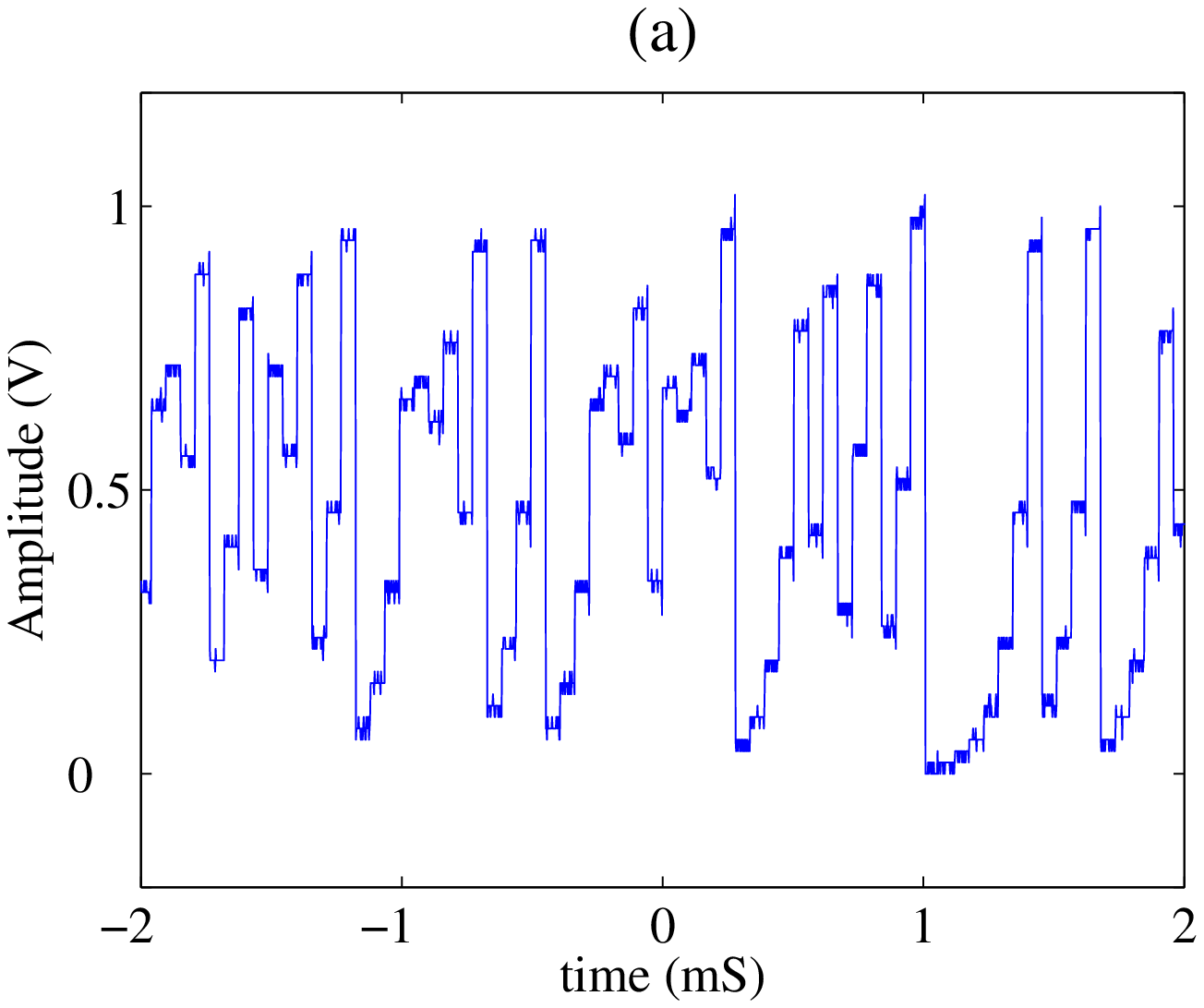} 
 \includegraphics[width= 9. cm, height=5.5 cm] {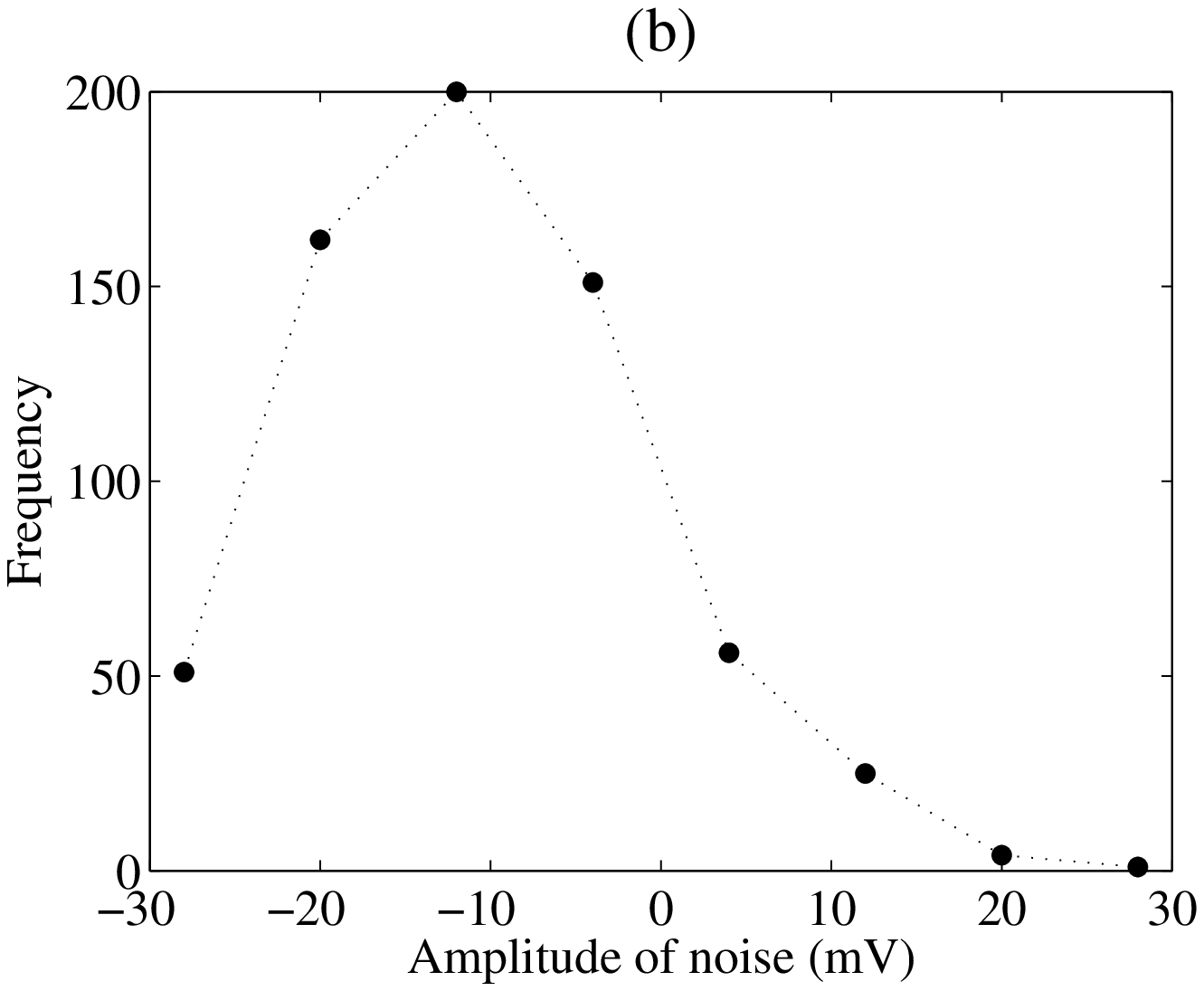}
\caption{\sl ({\rm a}) The time series with chaotic dynamics generated by the
tent map for $\mu=2$. ({\rm b}) The histogram of the noise estimated by means of equation (\ref{noise}).}
 \label{fig-SerieT}
 \end{figure}

In order to explore the full range of the
dynamics accessible to this circuit, we experimented with different
values of R$_5$ and R$_6$. These resistors were adjusted in the
closed interval [1 k$\Omega$, 4 k$\Omega$]. Then $\mu$ was varied to obtain the
bifurcation diagram shown in Fig.~\ref{figdatos}. In this figure,
fixed points, periodic oscillations, period-doubling cascade and
chaos can be clearly seen. From Fig.~\ref{figdatos}, it can be
seen that the circuit exhibits the entire range of behaviors of the
tent map. In fact, our experimental results of the dynamics of this
circuit are found to be in good agreement with the theoretical
values.
 Figure \ref{fig-SerieT} (a) shows a time series of the output
 voltage for $\mu=2$. In our measurements, each experimental time series contained 650
 points collected for different values of the bifurcation parameter $\mu$. Figure \ref{fig-SerieT} (b)
 shows the histogram of the noise calculated over the 650 points. The noise time series $r_n$ was estimated by the following equation

\begin{equation}\label{noise}
 r_{n} = x_{n+1}-f(x_{n})~,
 \end{equation}
where $x_n$ and $x_{n+1}$ are the experimental data of the tent map circuit, and $f(\cdot)$ is given by equation (\ref{ectent}).

%

\section{Conclusion}\label{S_Cls}
A very simple tent map electronic circuit has been presented here
and its implementation using only analog components as operational
amplifiers, diodes, and resistors was also provided. Therefore,
it can be assembled even by students at the level of an undergraduate laboratory.
Its experimental behavior was tested and compared with the numerical behavior given
by the tent map difference equation. The circuit that replicates the
whole known range of behaviors of the tent map has been determined.
The employed techniques are simple and the approach can be extended
to other types of maps such as the piecewise linear or piecewise
smooth maps. Such circuit realizations have many potential
applications, for example: random number generation,
frequency hopping, ranging, and spread-spectrum communications.
Finally, we notice that this design can be manufactured in just one
chip because the final electronic circuit contains only
semiconductors and passive components.

\subsection*{Acknowledgements}

This work was supported by FAI-UASLP 2007 under contract
C07-FAI-11-38.74 and also partially through the CONACyT project
46980.


 \end{document}